\documentclass[dvips]{article}
\usepackage{supertabular,lscape,epsfig}
\usepackage{amssymb}
\usepackage{amsmath}
 
\DeclareSymbolFont{ppa}{OT1}{ppl}{m}{it}
\DeclareMathSymbol{\vv}{\mathalpha}{ppa}{'166}
 
\begin{document}
 
\newcommand{\dd}{\,{\rm d}}
\newcommand{\ie}{{\it i.e.},\,}
\newcommand{\etal}{{\it et al.\ }}
\newcommand{\eg}{{\it e.g.},\,}
\newcommand{\cf}{{\it cf.\ }}
\newcommand{\vs}{{\it vs.\ }}
\newcommand{\zdot}{\makebox[0pt][l]{.}}
\newcommand{\up}[1]{\ifmmode^{\rm #1}\else$^{\rm #1}$\fi}
\newcommand{\dn}[1]{\ifmmode_{\rm #1}\else$_{\rm #1}$\fi}
\newcommand{\upd}{\up{d}}
\newcommand{\uph}{\up{h}}
\newcommand{\upm}{\up{m}}
\newcommand{\ups}{\up{s}}
\newcommand{\arcd}{\ifmmode^{\circ}\else$^{\circ}$\fi}
\newcommand{\arcm}{\ifmmode{'}\else$'$\fi}
\newcommand{\arcs}{\ifmmode{''}\else$''$\fi}
\newcommand{\MS}{{\rm M}\ifmmode_{\odot}\else$_{\odot}$\fi}
\newcommand{\RS}{{\rm R}\ifmmode_{\odot}\else$_{\odot}$\fi}
\newcommand{\LS}{{\rm L}\ifmmode_{\odot}\else$_{\odot}$\fi}
 
\newcommand{\Abstract}[2]{{\footnotesize\begin{center}ABSTRACT\end{center}

\vspace{1mm}\par#1\par
\noindent
{~}{\it #2}}}
 
\newcommand{\TabCap}[2]{\begin{center}\parbox[t]{#1}{\begin{center}
  \small {\spaceskip 2pt plus 1pt minus 1pt T a b l e}
  \refstepcounter{table}\thetable \\[2mm]
  \footnotesize #2 \end{center}}\end{center}}
 
\newcommand{\TableSep}[2]{\begin{table}[p]\vspace{#1}
\TabCap{#2}\end{table}}
 
\newcommand{\FigCap}[1]{\footnotesize\par\noindent Fig.\  %
  \refstepcounter{figure}\thefigure. #1\par}
 
\newcommand{\TableFont}{\footnotesize}
\newcommand{\TableFontIt}{\ttit}
\newcommand{\SetTableFont}[1]{\renewcommand{\TableFont}{#1}}
 
\newcommand{\MakeTable}[4]{\begin{table}[htb]\TabCap{#2}{#3}
  \begin{center} \TableFont \begin{tabular}{#1} #4
  \end{tabular}\end{center}\end{table}}
 
\newcommand{\MakeTableSep}[4]{\begin{table}[p]\TabCap{#2}{#3}
  \begin{center} \TableFont \begin{tabular}{#1} #4
  \end{tabular}\end{center}\end{table}}
 
\newenvironment{references}%
{
\footnotesize \frenchspacing
\renewcommand{\thesection}{}
\renewcommand{\in}{{\rm in }}
\renewcommand{\AA}{Astron.\ Astrophys.}
\newcommand{\AAS}{Astron.~Astrophys.~Suppl.~Ser.}
\newcommand{\ApJ}{Astrophys.\ J.}
\newcommand{\ApJS}{Astrophys.\ J.~Suppl.~Ser.}
\newcommand{\ApJL}{Astrophys.\ J.~Letters}
\newcommand{\AJ}{Astron.\ J.}
\newcommand{\IBVS}{IBVS}
\newcommand{\PASP}{P.A.S.P.}
\newcommand{\Acta}{Acta Astron.}
\newcommand{\MNRAS}{MNRAS}
\renewcommand{\and}{{\rm and }}
\section{{\rm REFERENCES}}
\sloppy \hyphenpenalty10000
\begin{list}{}{\leftmargin1cm\listparindent-1cm
\itemindent\listparindent\parsep0pt\itemsep0pt}}%
{\end{list}\vspace{2mm}}
 
\def\TYLDA{~}
\newlength{\DW}
\settowidth{\DW}{0}
\newcommand{\dw}{\hspace{\DW}}
 
\newcommand{\refitem}[5]{\item[]{#1} #2%
\def\REFARG{#3}\ifx\REFARG\TYLDA\else, {\it#3}\fi
\def\REFARG{#4}\ifx\REFARG\TYLDA\else, {\bf#4}\fi
\def\REFARG{#5}\ifx\REFARG\TYLDA\else, {#5}\fi.}
 
\newcommand{\Section}[1]{\section{#1}}
\newcommand{\Subsection}[1]{\subsection{#1}}
\newcommand{\Acknow}[1]{\par\vspace{5mm}{\bf Acknowledgements.} #1}
\pagestyle{myheadings}
 
\newfont{\bb}{ptmbi8t at 12pt}
\newcommand{\xrule}{\rule{0pt}{2.5ex}}
\newcommand{\xxrule}{\rule[-1.8ex]{0pt}{4.5ex}}
\def\thefootnote{\fnsymbol{footnote}}

\begin{center}
{\Large\bf The Optical Gravitational Lensing Experiment\footnote{Based on 
observations obtained with the 1.3-m Warsaw telescope at Las Campanas 
Observatory of the Carnegie Institution of Washington.}.\\
Difference Image Analysis of LMC and SMC Data. The Method} 
\vskip1cm
{\bf K. ~~\.Z~e~b~r~u~\'n$^{1,2}$,~~I. ~~S~o~s~z~y~\'n~s~k~i$^{1,2}$~~ and 
~~P.R. ~~W~o~\'z~n~i~a~k$^{2,3}$} 
\vskip3mm
{$^{1}$ Warsaw University Observatory, Al. Ujazdowskie 4, 00-478 Warsaw, Poland\\
e-mail: (zebrun,soszynsk)@astrouw.edu.pl\\
$^{2}$ Princeton University Observatory, Princeton, NJ 08544--1001, USA\\ 
$^{3}$ Los Alamos National Observatory, MS-D436, Los Alamos NM 85745, USA\\
e-mail: wozniak@lanl.gov}
\end{center}

\vspace{0.5cm}

\Abstract{We describe the Difference Image Analysis (DIA) algorithms and 
software used to analyze four years (1997--2000) of OGLE-II photometric 
monitoring of the Magellanic Clouds, the calibration, the photometric error 
analysis and the search for variable stars. A preliminary analysis of 
photometric errors is based on the field LMC$\_$SC2. A full catalog of more 
than 68~000 variable stars is presented in a separate 
publication.}{Techniques: photometric -- Methods: data analysis -- Magellanic 
Clouds} 
\vspace*{-9pt}
\Section{Introduction}
The Optical Gravitational Lensing Experiment -- OGLE (Udalski, Kubiak and 
Szyma\'nski 1997) is a long term observing project which the original goal was a 
search for dark matter in our Galaxy using microlensing phenomena (Paczy\'nski 
1986). During more than 8 years of project duration (OGLE-I: 1992--1995, and 
OGLE-II: 1997--2000) a huge number of images of the densest fields like the 
Galactic bulge, Magellanic Clouds, Galactic disk were collected. A modified 
{\sc DoPhot} package (Schechter, Mateo and Saha 1993) was used to derive 
photometry for millions of stars. The advantages of {\sc DoPhot} are the speed 
and high efficiency for dense star fields, such as those observed by the OGLE 
project. 

Most fields observed by OGLE are very crowded. In such fields very often more 
than one star contributes to the total light within a single observed Point 
Spread Function (PSF). In many cases variability refers to one of several 
components of the PSF, \ie only some part of the total measured flux is due to 
the variable star. Therefore, the photometry of a variable star is often 
biased by a blending star(s). 

\hglue-3pt Image subtraction can solve many of the problems described above. 
The method was introduced in the 90's. Two algorithms have been successfully 
applied: Fourier division (Tomaney and Crots 1996, Alcock \etal 1999) and a 
linear kernel decomposition in real space (Alard and Lupton 1998, Alard 2000). 
The latter algorithm was also implemented by Wo\'zniak (2000, hereafter Paper~I) 
in the Difference Image Analysis (DIA) package. In this paper we will use the 
software and nomenclature introduced in Paper~I. We describe our modifications 
of the DIA technique, as applied to the 11 SMC and 21 LMC fields observed by 
OGLE-II in 1997--2000. The full catalog of variable stars in all these 
fields is presented in another paper (\.Zebru\'n \etal 2001). 

\Section{Observational Data}
\hglue-3pt We used the data obtained during four observing seasons of the 
OGLE-II project: from January 1997 until May 2000. The data were collected 
with the 1.3-m Warsaw telescope at the Las Campanas Observatory, which is 
operated by the Carnegie Institution of Washington. The telescope was equipped 
with the ``first generation'' camera with a SITe ${2048\times2048}$ CCD 
detector. The pixel size was 24~$\mu$m giving the 0.417 arcsec/pixel scale. 
The observations were conducted in the drift-scan "slow-mode" with the gain 
3.8e$^-$/ADU and readout noise 5.4e$^-$. Single frame's size is 
${2048\times8192}$ pixels and corresponds to $14\zdot\arcm2\times57\arcm$ in 
the sky. The details of the instrumentation setup can be found in Udalski, 
Kubiak and Szyma\'nski (1997). 

More than 60 fields were observed in the Magellanic Clouds. Each of them 
covered approximately 0.22 square degrees in the sky. We analyzed only 
frequently observed fields, 21 in the LMC and 11 in the SMC. About 400 {\it 
I}-band observations were collected for each of the LMC fields, and about 300 
{\it I}-band observations for each of the SMC fields. For each of the LMC and 
SMC fields about 40 and 30 observations in the {\it V} and {\it B}-band, 
respectively, were also collected. The effective exposure time was 237, 173 
and 125 seconds for {\it B}, {\it V} and {\it I}-band, respectively. Mean 
seeing of the entire data set was about 1\zdot\arcs34. 

During the DIA analysis we used all collected {\it I}-band observations, 
regardless of seeing. The data were flat-fielded by the standard OGLE 
procedure (Udalski, Kubiak and Szyma\'nski 1997). Uncompressed raw {\it I}-band 
images filled almost 400~GB of disk space. 

On the reference frames (see Section~3) we detected a total of about 
$2\times10^7$ objects. The total number of photometric measurements for these 
objects was over $6\times10^9$. The DIA analysis provided over $8\times10^4$ 
candidates for variable stars. After removing various artifacts (see 
Section~7) we were left with about $7\times10^4$ candidates. All these stars 
are presented in the catalog (\.Zebru\'n \etal 2001). 

\Section{Data Analysis}
We used the DIA data pipeline that is based on the recently developed image 
subtraction algorithm described by Alard and Lupton (1998) and Alard (2000). 
The software package that uses this method and pipeline scheme was developed 
by Wo\'zniak (2000) and is described in Paper~I. 

Following Paper~I, a special attention was paid to the selection of 
frames for preparation of the reference image. We selected frames with the 
best seeing, small relative shifts, low background, and free of satellite 
trails and other artifacts. Most cosmic rays were removed in the process of 
averaging of the reference image. In Section~4 we describe some modifications 
to that part of the original package. A weighted average of 20 best frames was 
adopted as the reference image. This image is used to obtain the DC signal, 
also known as the reference flux, for the DIA photometry. The reference frame 
has the same coordinate grid as the OGLE template for easy comparisons between 
standard OGLE and DIA databases. We also kept the original partitioning into 
$4\times64$ subframes, $512\times128$ pixels each. 

For the actual difference photometry we adopted the script {\sc Pipe} as 
presented in Paper~I. A long list of input parameters required only minor 
adjustments between the Galactic bulge and the Magellanic Clouds data. Also 
the search for variables is basically the one from Paper~I, except for the 
removal of the most obvious artifacts (see Section~7). 

We found that the number of detected variables is very sensitive to one of the 
pipeline parameters, namely CORR-PSF, which sets the minimum required value 
for the correlation coefficient with the PSF for a candidate group of variable 
pixels. For details see Section~3.9 of Paper~I. Lower values obviously 
generate more candidate variable objects. Changing CORR-PSF from 0.7 to 0.6 
will result in about 30\% longer list of objects, unfortunately with 
increasing proportion of artifacts. In Section~7 we briefly describe problems 
of selecting the optimal parameter values. All results presented in the 
catalog (\.Zebru\'n \etal 2001) were obtained with CORR-PSF of 0.7, with the 
exception of LMC$\_$SC2, where experimentally we used CORR-PSF=0.6. 

The tools for reliable DC photometry are still not included with any of the 
DIA software packages. Although we describe here the DIA results, some 
measurements, especially for calibration purposes, had to be made with profile 
photometry package like {\sc DoPhot}. We compare our results with OGLE data, 
which were also processed with the {\sc DoPhot} package. To avoid confusion, 
we list below some introduced abbreviations. 
\vspace*{-6pt}
\begin{itemize}
\itemsep=-4pt
\item OGLE {\sc DoPhot} -- refers to the results with the standard {\sc DoPhot}
pipeline in OGLE, as described by Udalski, Kubiak and Szyma\'nski (1997),
\item DIA {\sc DoPhot} -- results from reference frames created by
{\sc Make$\_$template} script of the DIA package processed with {\sc DoPhot},
\item DIA photometry (DIA) -- difference photometry (AC) and simple
DC photometry on the reference image obtained with the DIA package.
\end{itemize}

The resulting photometric data from DIA are expressed in linear flux units. 
The transformation to magnitude scale is described in Section~5. The DC flux 
was measured on a reference image independently with DIA package and {\sc 
DoPhot} package. The details of selection of the adopted value of DC signal 
for the measured star are described in Section~5. 

\section{Changes to the DIA Package}
\subsection{Cosmic Ray Killer (CRK)}
Magellanic Cloud images are exposed by about 30\% longer than those of the 
Galactic bulge observations and, as a result, the number of cosmic rays on 
reference images is larger. The total number of cosmic ray hits accumulated on 
20 frames stacked to form a reference image can pose a significant problem for 
photometric processing. To remove cosmic rays from reference frame we created 
a special procedure, attached to the program {\sc mstack} (\cf Paper~I, 
Section~3.6).

We take a series of resampled frames that will be coadded to generate 
reference frame and analyze pixels with the same $X,Y$ coordinates. We 
calculate the median and $\sigma$ (standard deviation) of all flux values to 
be stacked. Then we check whether the value of flux in a brightest pixel that 
is coadded deviates from the median by more than $50\sigma$. Such a pixel is a 
candidate for rejection by CRK. 

For these candidate pixels we examine the median of 8 neighboring pixels. If 
the pixel in question is at least twice as bright as the median of its 
neighbors, it is marked as a cosmic ray. 

The pixels marked as cosmic rays are not used to calculate the average value 
of a pixel on a reference frame by {\sc mstack} program. 

The effect of Cosmic Ray Killer application is shown in Fig.~1. Both images 
show the same subframe of the LMC$\_$SC2 field. The left image was obtained 
without using CRK. Cosmic rays are rather easy to spot. One of them appears as 
a short line. The others are considerably brighter than their vicinity. The 
right image has the cosmic rays removed. 
\begin{figure}[htb]
\centerline{\includegraphics[width=6cm]{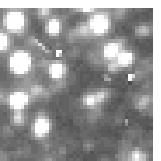}
\hskip5mm\includegraphics[width=6cm]{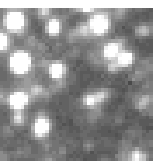}}
\FigCap{Removal of cosmic rays. The same section of the reference image for 
LMC$\_$SC2 field processed by standard {\sc mstack} program (left) and our 
modified version (right) is shown. All clearly detectable cosmic rays are 
cleanly removed.} 
\end{figure}

\subsection{Removal of Bad Column}
The CCD used by OGLE-II contained a group of bad columns. On a drift-scan they 
formed a line-like feature along a whole scan with a thickness of several 
pixels: a BAD COLUMN. 

Reduction of the image includes the transformation onto a uniform coordinate 
grid with a bicubic spline function (\cf Paper~I, Section~3.5). Near the BAD 
COLUMN the function strongly oscillates because it meets a group of pixels of 
the same large value. This results in incorrect values of flux on the 
transformed picture. 

To remove this effect the following procedure was introduced and added to the 
{\sc resample} program. Before applying bicubic spline interpolation we 
localize BAD COLUMN on the frame. Then we replace the value of the each pixel 
in the BAD COLUMN with an average frame background value with a random 
Gaussian noise superimposed. In the next step we perform a bicubic spline 
interpolation. Finally, we restore BAD COLUMN on a resampled frame by taking 
advantage of imperfect pointing and renormalizing pixel by pixel averages. 
Unusable values are marked as saturated and ignored later. 

We also improved the quality of the results by introducing a new {\sc sfind}, 
a star finding procedure. A better handling of very bright and saturated stars 
near the center of analyzed subframe eliminated wrong background estimates. 
The original version of {\sc sfind} (Paper~I, Section~3.3) calculated 
background value in an annulus near the center of the frame. This background 
was used later as a threshold for star detection. This works well when there 
are no very bright and saturated stars on the subframe. However bright stars 
are common in the Magellanic Clouds. Occasionally the original {\sc sfind} 
failed because background was calculated from saturated pixels. To avoid this 
problem we used the median of background estimates in many locations. 

\Section{Transformation to the Magnitude System}
The Difference Image Analysis provides flux differences for variable stars. In 
order to convert these into more familiar magnitudes it is necessary to 
determine the baseline flux level (the zero point, or the DC flux), and the 
relation between the flux differences in DIA photometry and in DIA {\sc 
DoPhot} photometry. As the first step to establish this relation, we 
determined the photometry of stars on the reference frames of all fields with 
DIA {\sc DoPhot}. As each DIA reference image was obtained by co-adding the 
best 20 frames for each field, it was possible to make more accurate 
photometry and to reach deeper than in a single template frame used by the 
OGLE-II for on-line data processing: typically the number of detected stars 
almost doubled. The zero point for the magnitude scale on the reference image 
was determined by comparing stars brighter than 17.5~mag with the OGLE 
photometric databases (Udalski \etal 1998, 2000). The zero point was obtained 
separately for each of 256 segments of a reference frame. 

At this point we also checked the linearity of DIA {\sc DoPhot} measurements 
of reference images. The differences between mean OGLE {\sc DoPhot} magnitudes 
of stars and DIA {\sc DoPhot} magnitudes of stars for LMC$\_$SC2 are shown in 
Fig.~2 as a function of star brightness. Only the stars for which the 
cross-identification was better than 0.1 pixel were taken into account. It is 
clearly seen that co-adding twenty frames does not spoil linearity. 
\begin{figure}[htb]
\centerline{\includegraphics[width=10.7cm]{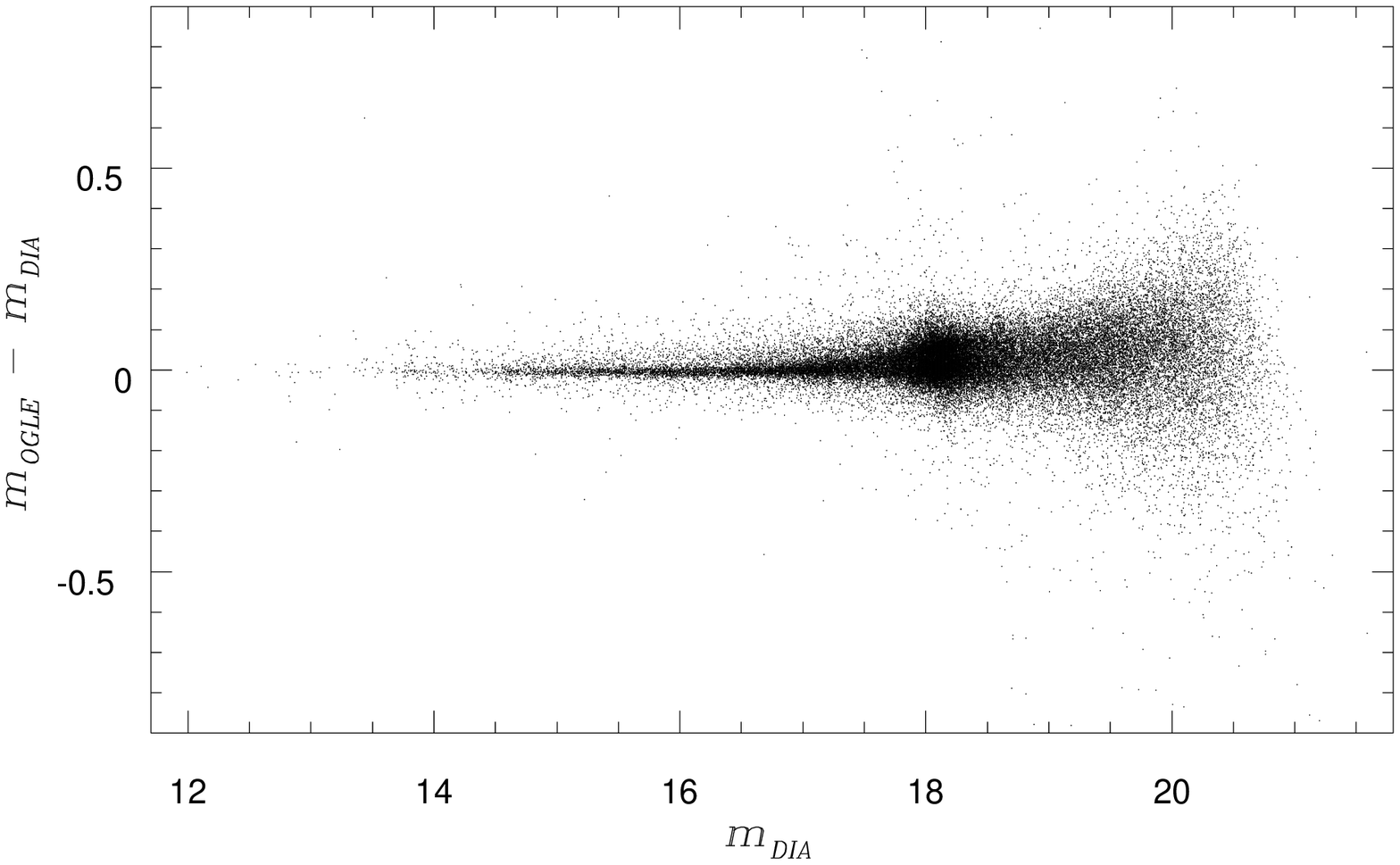}}
\FigCap{Offset between the {\sc DoPhot} magnitude measured on the reference 
frame ($m_{\rm DIA}$) and the magnitude in the OGLE database ($m_{\rm OGLE}$) 
as a function of $m_{\rm DIA}$ for 50000 stars from LMC\_SC2. One can see good 
agreement between OGLE and DIA for whole range of photometry.} 
\end{figure}

The brightness, in magnitudes, of each photometric point was calculated using 
formula: 
$$m_i=-2.5\cdot\log(f_{\rm DC}+a\cdot f_{\rm ACi})+{\rm zero\_point}\eqno(1)$$
where:
$f_{\rm DC}=10^{-0.4\times m_{\rm DC}}$ -- brightness (in flux) on the 
reference frame, expressed in {\sc DoPhot} flux units,\\ 
$f_{ACi}$ -- difference brightness from DIA photometry,\\
$a$ -- coefficient scaling DIA flux to flux measured by {\sc DoPhot},\\
$\rm zero\_point$ -- magnitudes zero point taken from 
comparison with OGLE database.

To determine the coefficient $a$ we used two methods. The first method was 
described in Paper~I. We used DIA photometry and DIA {\sc DoPhot} photometry 
obtained on a reference frame. For isolated stars the contribution of 
neighboring stars to the total flux is less than 1\% and can be neglected. 
\begin{figure}[htb]
\centerline{\includegraphics[width=10.7cm]{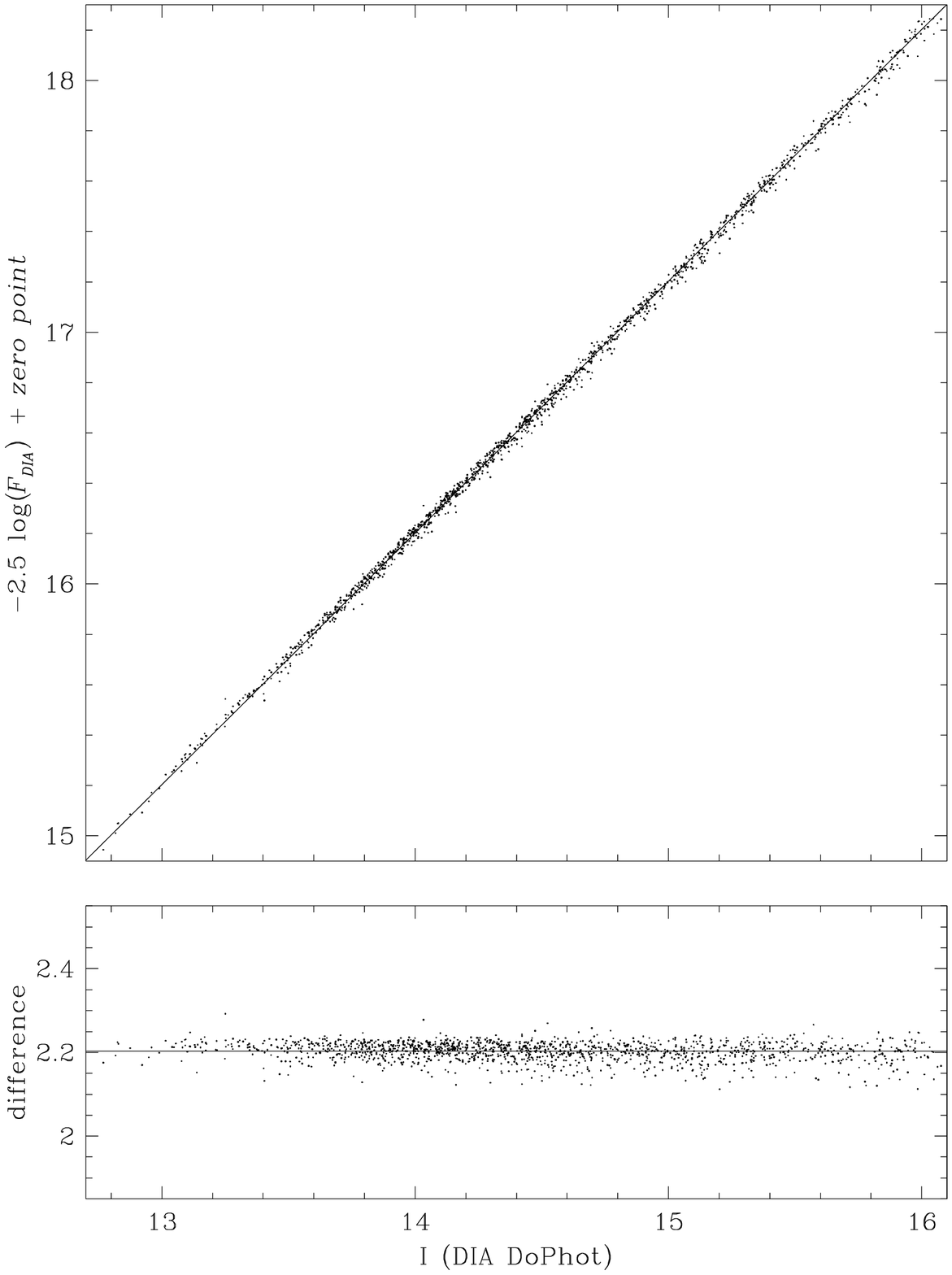}}
\FigCap{First method of finding the transformation coefficient between DIA 
fluxes and magnitudes. {\it I}-band magnitudes measured by DIA {\sc DoPhot} 
are plotted against the ${-2.5\times\log}$ of the DIA DC flux (upper panel). 
The difference between these two magnitudes (lower panel) defines the 
transformation factor. The best fit gives the transformation factor of 
${7.61\pm0.06}$.} 
\end{figure}

We expressed the brightness of stars in magnitude units. The upper panel of 
Fig.~3 shows the relationship between {\it I}-band magnitude obtained by 
DIA {\sc DoPhot} and ${-2.5\times}$ logarithm of DIA DC flux. The relation is 
obviously linear. We used the least squares method to fit a linear function. 
The slope is ${1.004{\pm}0.005}$. We derived the $a$ coefficient of Eq.~(1) from 
the magnitude difference (lower panel of Fig.~3) between $x$ and $y$ axis in 
the upper panel of the Fig.~3 using the relation 
$$a=10^{\mbox{difference/2.5}}.\eqno(2)$$
The line in the lower panel of Fig.~3 is the best linear fit of the magnitude 
difference yielding the value of ${a=7.61\pm0.06}$. 

In the second method the coefficient $a$ was obtained by comparing light 
curves of Cepheids detected in the DIA data and in the OGLE database. These 
stars are bright, photometry has good quality and more importantly the catalog 
of Cepheids has already been published by OGLE (Udalski \etal 1999a, 1999b).  
For each Cepheid we calculated the coefficient $a$ by minimizing the 
differences between measurements for the same objects in both catalogs. In 
Fig.~4 we present individual estimates of the coefficient for each Cepheid. 
The median is ${a=7.60\pm0.08}$, making both results consistent and in 
very good agreement with the value found in Paper~I for the OGLE Galactic 
Bulge data. Field to field variations were within the 1\% error given above. 
\begin{figure}[htb]
\centerline{\includegraphics[width=10.7cm]{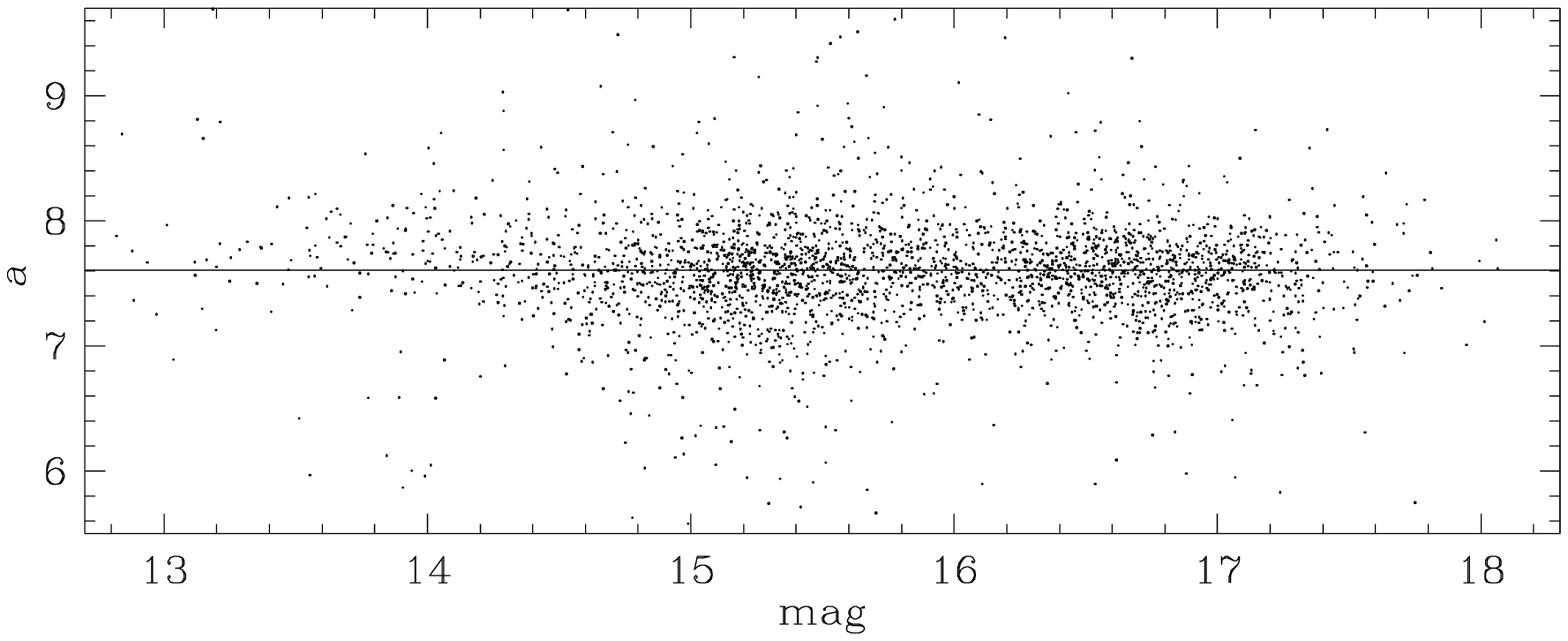}}
\FigCap{Transformation coefficient from DIA flux to magnitude units calculated 
with Cepheid variable stars. The line corresponds to our adopted value, 
${7.60\pm0.08}$.} 
\end{figure}

As we mentioned before, $f_{DC}$ is the DC signal flux expressed in {\sc 
DoPhot} units. We emphasize here that the DC flux level is difficult to 
determine for any variable located in a crowded field, no matter which method 
is used for photometry. As long as there is a high probability that any given 
PSF is composed of a blend of several stars, with only one of them variable, 
there is no general way to figure out what is the DC component corresponding 
to the variable object. One of the few exceptions is provided by gravitational 
microlensing, which can be modeled as a purely geometrical effect. Another 
example is a light curve of a detached eclipsing binary, for which the 
contribution of the "third light" can be determined. But there is no general 
solution. The virtue of the DIA is that it presents this generic problem with 
full clarity, while this issue is not so obvious if a software like {\sc 
DoPhot} is used. However, the problem is always there, hidden or not. 

The DC flux measurement on a reference image from the DIA package is not very 
precise. This is because DIA is not modeling the star's vicinity on a 
reference frame and not removing nearby stars prior to calculating the flux. 
To minimize the problem of proper DC signal calculation we also calculated a 
DIA {\sc DoPhot} photometry on a reference frame, which was converted to DC 
flux for each star detected on the reference frame using relationships derived 
above. At this point the best solution was to adopt DIA {\sc DoPhot} flux as 
the correct DC signal for a variable star. However there are some caveats of 
this procedure. For example small differences in the DC flux, after 
transformation to magnitude system, result in very large differences. Fig.~5 
shows two light curves of a single star from LMC\_SC4 field. For these light 
curves we used two DC flux values that differ by 0.5~mag. It is clear that the 
light curves are not the same. The differences between both light curves are 
even as large as 6~mag. 
\begin{figure}[htb]
\centerline{\includegraphics[width=10.7cm]{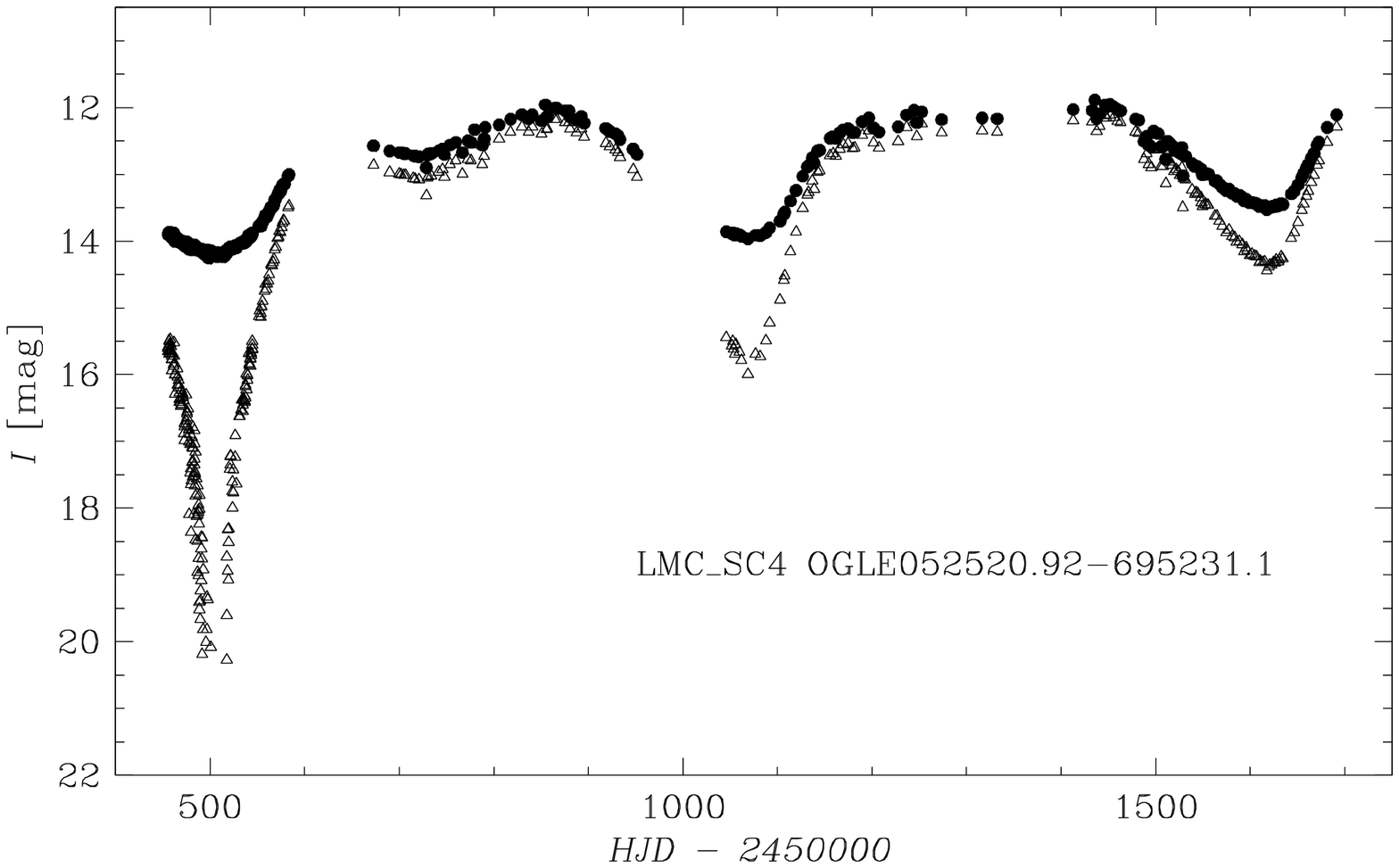}}
\FigCap{Light curve of a variable star OGLE052520.92-695231.1. The difference 
of about 0.5~mag in DC magnitude radically changed the shape of the light 
curve. Full circles indicate magnitudes computed with ${m_{\rm 
DC}\approx13.2}$, and empty triangles indicate magnitudes obtained using 
${m_{\rm DC}\approx13.7}$. The problem of wrong DC measurements applies to 
about 5\% of all variable objects.} 
\end{figure}

Such wrong DC flux estimates are caused by different break up of some blends 
in separate runs of {\sc DoPhot} and occurred in our database for about 5\% of 
all variables. We used the following procedure to correct DC flux. After 
searching the DIA database for objects that could be identified on OGLE 
template to better than 0.3 pixels, we compared OGLE {\sc DoPhot} and DIA 
photometry. When the median of differences between OGLE {\sc DoPhot} and DIA 
measurements was larger than 0.1~mag we adopted the more noisy but less biased 
DC signal from the DIA photometry. 

The DIA AC signal is measured on subtracted images at the position of the 
variable object (see Paper~I, Section~3.9). Of course the variable may be 
blended with a brighter star which is not resolved by {\sc DoPhot}. Therefore 
the variable stars are not always identified positionally with a high 
precision on the reference frame. We decided to set a limit of identification 
distance between the position of a variable star and the position of the 
nearest object in the DIA {\sc DoPhot} list of stars. When the distance was 
smaller than the limit we used the DC signal from the DIA {\sc DoPhot} 
photometry, otherwise we adopted as the DC signal the flux from DIA photometry 
on the reference image. The separation limit depended linearly on the 
magnitude, ranging from 4 pixels for the brightest stars to 1 pixel at the 
faint end. Occasionally, there was no detectable {\sc DoPhot} star on the 
reference frame at the location of a variable star, which usually means that 
the variable was above the detection threshold on some frames only, and that 
none of those contributed to the reference image. Some other cases were likely 
pseudo-variables, due to random increase of noise on some subtracted frames. 
In some cases we found more than one star to be closer than the adopted 
separation limit. In these cases we used the DC signal for the brightest of 
those stars. 

\Section{Noise Properties}
The DIA delivers very precise measurements of the AC signal. Therefore, light 
curves of variable stars are smoother than those from {\sc DoPhot} data. In 
this Section we assess the noise characteristics for our DIA data. 

The DIA errors are affected by photon noise, but obviously they are larger. 
Wo\'zniak (2000) used data for a few hundred bulge microlensing events to 
estimate these errors. Microlensing is very rare in the LMC and SMC, but a 
very similar procedure can be applied to constant stars. Wo\'zniak (2000) 
fitted a point model to the microlensing event light curve and calculated 
standard deviation of measurements around the model in the unmagnified region, 
where the flux is practically constant. As most stars are constant and trivial 
to model, they are well suited for DIA error estimates. However this approach 
required slight modifications of the pipeline, by basically bypassing {\sc 
getvar}, which normally only generates the list of positions for candidate 
variables. For noise estimates on a given frame we made the list of all 
detectable stars with no variables closer than 5 pixels, and supplied this 
directly to the {\sc phot} procedure, which extracts the actual photometry by 
filtering through the PSF. 

The image subtraction software is very CPU intensive, and for the first 
edition of the LMC and SMC variable star catalog we made only a restricted 
error analysis, based on one field only, LMC$\_$SC2. In this field we detected 
about 6200 variable star candidates out of about 730~000 stars total (\.Zebru\'n 
\etal 2001). The time to perform the DIA photometry is proportional to the 
number of measured stars. To save the CPU time we selected eight $512 \times 
128$ pixel sub-frames uniformly distributed throughout the LMC$\_$SC2 field. 

The following figures are based on the results from these subframes and 
include the data from about $10^5$ constant star light curves. The stars were 
grouped into 0.5~mag bins according to the DIA {\sc DoPhot} flux. In Fig.~6 we 
present the {\it rms} residuals around the median, normalized by photon noise 
(upper panel) and expressed in magnitudes $\sigma_I$ (lower panel). The best 
fit to the data is shown as solid line and described by: 
$$\frac{\sigma_F}{\sigma_{\rm ph}}=1.1671\times(0.8+2.736\times10^{-4}
\times F)^{1/2}.\eqno(3)$$
For this fit we did not use the brightest bin where photon noise is very low 
and imperfections of the PSF determination dominate the residuals. 
\begin{figure}[htb]
\centerline{\includegraphics[width=9cm]{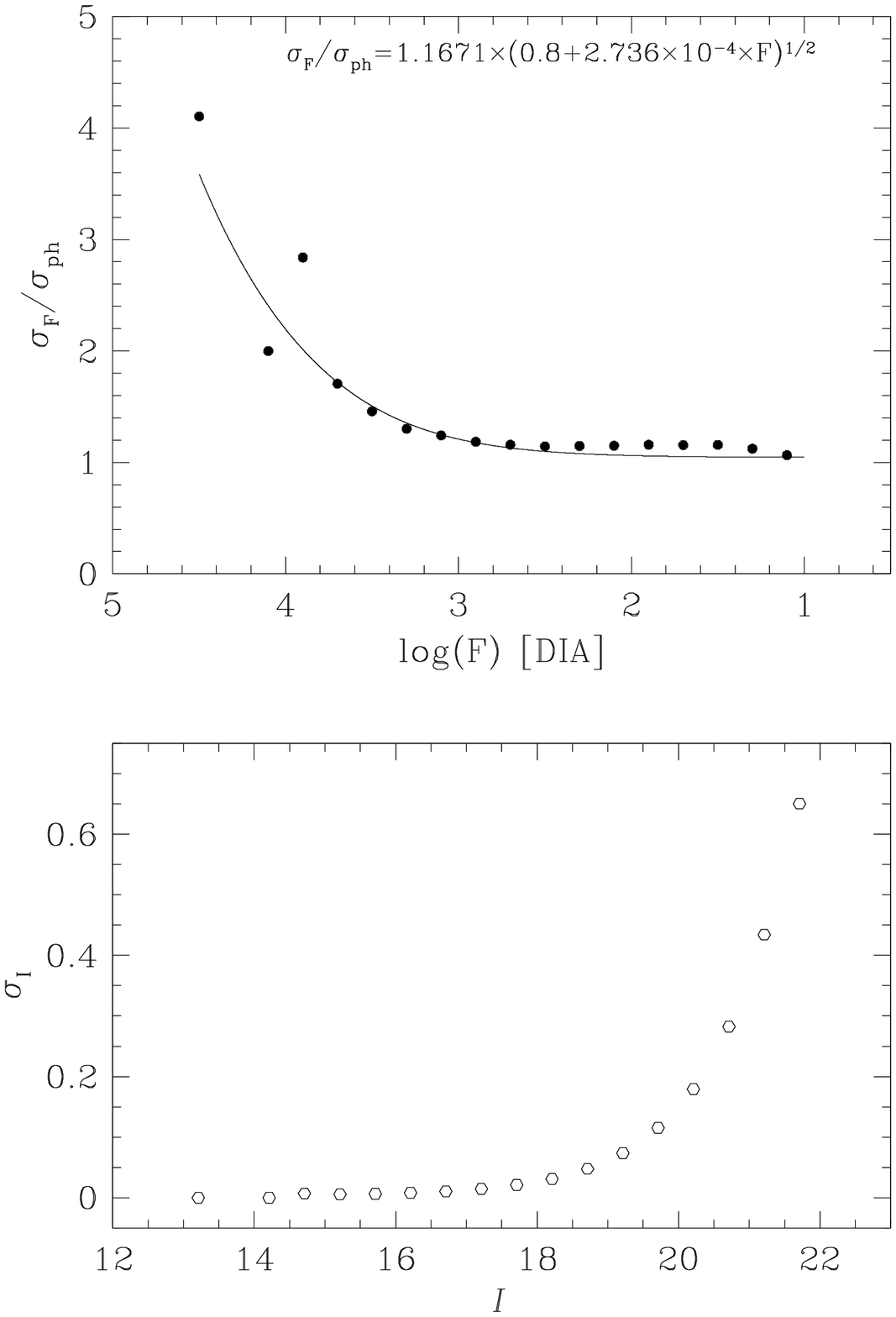}}
\FigCap{In the upper panel we plot the {\it rms} scatter around the median 
divided by the photon noise estimate as a function of flux logarithm for a set 
of 32~000 constant stars from LMC$\_$SC2 field. The flux bins correspond to 
0.5~mag. The solid line is the best fit to the data points (Eq.~3). 
{\it I}-band magnitude error for a range of magnitudes is given in the lower 
panel. The open circles correspond to the points in the upper panel. The 
errors for the brightest stars are at the level of 0.005~mag.} 
\end{figure}

The run of $\sigma_I$ is almost linear down to 18~mag. For the brightest stars 
($I<16$ mag) the errors are at the level of 0.005~mag. Then the errors grow to 
about 0.08~mag for stars of 19~mag and 0.3~mag for stars at ${I=20.5}$~mag 
(OGLE photometry limit, see Fig.~2). For the transformation of the DIA fluxes 
(ordinate of the filled points in the upper panel) to magnitudes (ordinate of 
the open circles in the lower panel) we used Eq.~(1). 

Fig.~7 shows the distribution of residuals for selected magnitude bins. The 
adopted error bars were normalized using Eq.~(3). The solid lines are 
Gaussian fits to the data. Only 1\% of all measurements belong to the 
non-Gaussian wings, except for the very brightest stars, for which the wings 
are significantly larger, again, due to imperfect modeling of the PSF. 
\begin{figure}[htb]
\centerline{\includegraphics[width=10.7cm]{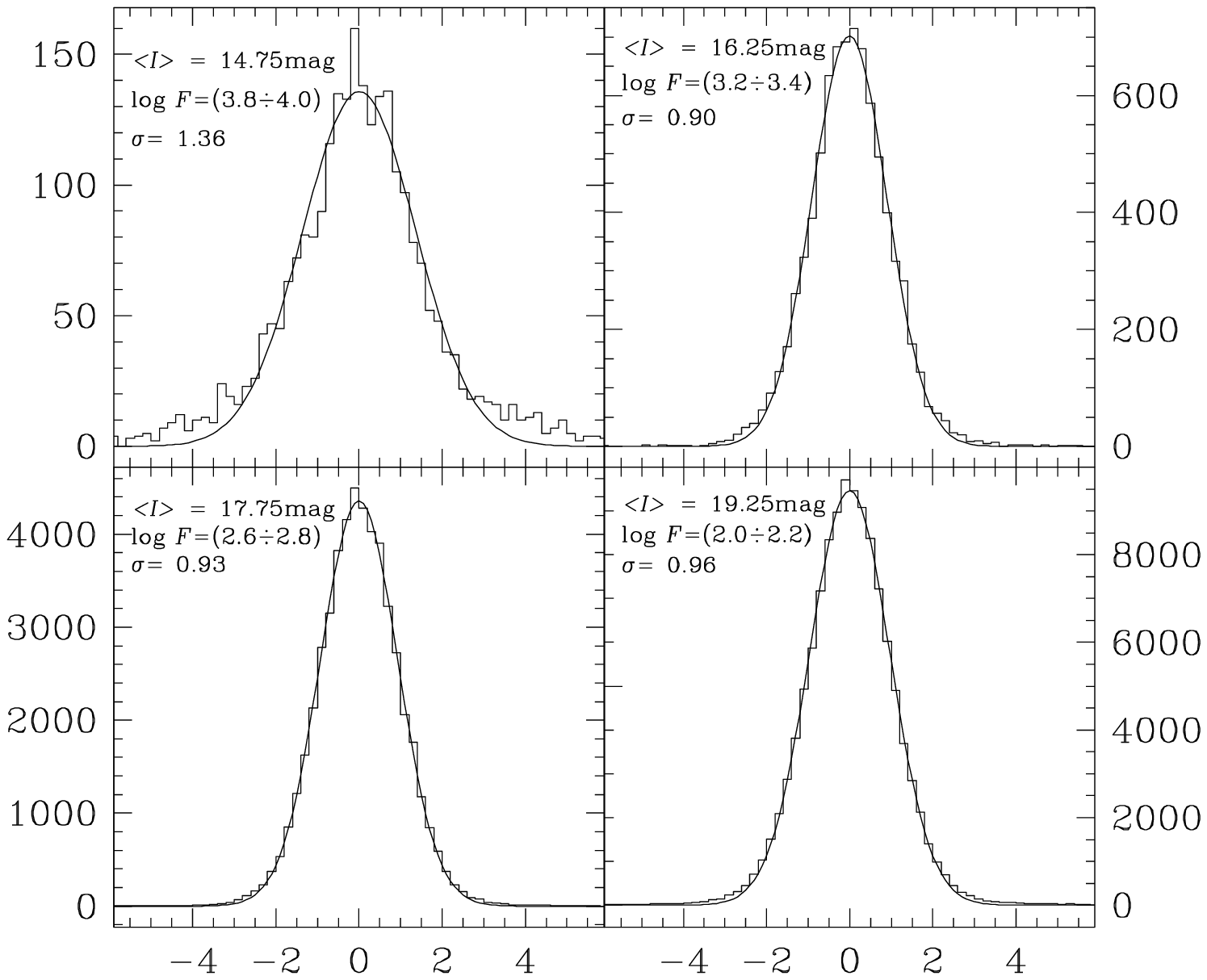}}
\FigCap{Histograms of normalized residuals around median flux in four selected 
magnitude bins. As the normalization factor we used the photon noise corrected 
by the fit from Eq.~(3). The error distribution is well approximated by a 
Gaussian fit (the solid lines), with the exception of the very brightest 
stars, for which systematics of the PSF model are noticeable.} 
\end{figure}

The error bars quoted in the catalog (\.Zebru\'n \etal 2001) are given by the 
photon noise corrected with Eq.~(3). The current error analysis is based 
on a single field, LMC\_SC2, and as such should be considered preliminary. 

\Section{The Results -- Variable Objects} 
While selecting about $10^5$ variables out of about $10^7$ stars, with a 
typical number of about 400 photometric data points, there is no way to 
complete the process manually. The only option is a fully algorithmic 
procedure, implementing a set of filters for selecting variables and rejecting 
artifacts. The initial algorithm to select variables is included in program 
{\sc getvar} in the DIA package (Paper~I, Section~3.9). In practice there is 
always a trade-off between the number of admitted artifacts and missed 
variables. At this time we are not able to present the optimum algorithm. Most 
likely the optimum selection process will have to be found empirically, as it 
requires a very complex multi parametric optimization which takes a lot of CPU 
time. Nevertheless we should be getting closer to the goal through the 
subsequent releases of the catalog. 

The present catalog basically adopts the algorithm as used in the analysis of 
the OGLE-II BUL\_SC1 field (Paper~I). The photometry for all objects from the 
DIA database of variable objects expressed in linear flux units as well as in 
magnitudes is presented in \.Zebru\'n \etal (2001). The catalog contains many 
known kinds of variables, \ie pulsators, eclipsing binaries etc, but it may 
still contain some artifacts. 

\subsection{The Artifacts}
Constant stars located close to a bright star account for a significant number 
of artifacts. Their pseudo-variability comes from the background variations in 
the wings of the bright star (variable or not, because atmospheric and 
instrumental scattering may also induce variable wings). 

To remove these artifacts we proceeded as follows. For each variable objects 
we calculated a cross-correlation function of light curves with every object 
within a 15 pixels radius. When the correlation between any two light curves 
was higher than a given threshold (0.7 in our case) we assumed that both stars 
varied in the same way. Next, we sorted the stars with similar light curves 
according to their brightness and identification with DIA {\sc DoPhot} star on 
reference frame. The brightest star was marked as the true variable, and all 
remaining stars were treated as artifacts. 

During the third observing season of OGLE-II the telescope mirror was 
realuminized. This had an effect on some faint stars which were close to 
bright stars. Some of these faint stars exhibited a sudden variability 
coincident in time with the aluminization. To remove these pseudo-variables 
from our sample we used a cross-correlation function again. First, we selected 
one star with such behavior in each of our fields as a template of a pseudo 
light curve. Next, we cross-correlated this light curve with the light curves 
of all stars in a given field. When any of the stars had cross correlation 
with our template star close to 1 or $-$1 we added this star to the list of 
suspected artifacts. Finally, we inspected by eye all these light curves and 
rejected many of them as artifacts. 

\subsection{Stars with High Proper Motion}
Eyer and Wo\'zniak (2001) recognized that pairs of variables from the catalog 
published by Wo\'zniak (2000) separated by about 3 pixels were in fact single 
non-variable stars with a detectable proper motion. One component of a pair 
showed monotonic increase in brightness, while the other varied in opposite 
sense, with the total flux roughly constant. Hence, the two light curves had 
the cross correlation coefficient close to ${-1}$. We found about 1000 such 
pairs in the LMC and about 300 in the SMC and excluded them from the catalog 
of variables (\.Zebru\'n \etal 2001). Their catalog will be presented in a 
separate paper. 

\Section{Summary}
We described application of the DIA method to the LMC and SMC observational 
data. We followed very closely the procedure developed in Paper~I, with minor 
modifications. The catalog of variables (\.Zebru\'n \etal 2001) should be 
considered to be the first, preliminary edition, to be gradually improved. We 
encourage the users to bring to our attention any problems they might 
encounter while using it. 

OGLE-III is using a new CCD camera with $15~\mu$m pixels, corresponding to 
0.25 arcsec/pixel and guided exposures. At better spatial resolution than 
available in OGLE-II, we are able to extract more precise signal with both 
{\sc DoPhot} and DIA photometry. The OGLE-II fields in the LMC and SMC are 
still observed with increased frequency. Therefore, we shall be able to obtain 
much better reference images in the near future and improve determinations of 
the DC flux. Higher quality light curves of all variables will follow.

\Acknow{We are indebted to Dr.\ B.\ Paczy\'nski for many valuable discussions 
and support during preparation of this paper. It is a pleasure to acknowledge 
that this work begun when two of us (KZ, IS) were visiting Department of 
Astrophysical Sciences at Princeton University, and one of us (PW) was a 
graduate student at that department. We would like to thank Dr.\ A.\ Udalski 
for numerous comments that substantially improved the paper. This work was 
partly supported by the KBN grant 5P03D 025 20 for I.\ Soszy\'nski and 5P03D 027 
20 for K.\ \.Zebru\'n. Partial support was also provided by the NSF grant 
AST-9830314 to B.\ Paczy\'nski.}

\end{document}